\newcommand{\ASPC}{   {\it Astron. Soc. Pac. Conf. Series} }
\newcommand{\RSPT}{   {\it Royal Society of London Philosophical Transactions Series A} }

\documentclass[referee]{raa}            

\usepackage{graphicx,times}             
\usepackage{enumerate}

\begin{document}

\title{A study of the relation between intensity oscillations and magnetic field parameters in a sunspot: Hinode observations}

   \volnopage{Vol.0 (200x) No.0, 000--000}      
   \setcounter{page}{1}          

   \author{A. Raja Bayanna
      \inst{1}
   \and Shibu K Mathew
      \inst{1}
   \and Brajesh Kumar
      \inst{1}
   \and Rohan E. Louis
      \inst{2}
   \and P. Venkatakrishnan
      \inst{1}
   }

   \institute{Udaipur Solar Observatory, Physical Research Laboratory, Dewali, Badi Road,
Udaipur 313 004, India {\it bayanna@prl.res.in}\\
	\and
Leibniz-Institut f\"{u}r Astrophysik Potsdam (AIP), An der Sternwarte 16, 14482, Potsdam, Germany\\
   }

   \date{Received~~2009 month day; accepted~~2009~~month day}

\abstract{ 
We present properties of intensity oscillations of a sunspot in the photosphere and chromosphere using G~band and Ca~{\sc ii}~H filtergrams from {\it Hinode}. Intensity power maps as function of magnetic field strength and frequency reveal reduction of power in G~band with increase in photospheric magnetic field strength at all frequencies. In Ca~{\sc ii}~H, however, stronger fields exhibit more power at high frequencies particularly in the 4.5~mHz--8.0~mHz band. Power distribution in different locations of the active region show that the oscillations in Ca~{\sc ii}~H  exhibit more power in comparison to that of G~band. We also relate the power in intensity oscillations with different components of the photospheric vector magnetic field using near simultaneous spectro-polarimetric observations of the sunspot from the {\it Hinode} spectropolarimeter. The photospheric umbral power is strongly anti-correlated with the magnetic field strength and its the line-of-sight component while there is a good correlation with the transverse component. A reversal of this trend is observed in the chromosphere with the exception at low frequencies ($\nu\le$~1.5 mHz). The power in sunspot penumbrae is anti-correlated with the magnetic field parameters at all frequencies (1.0~$\le\nu\le$~8.0~mHz) in both the photosphere and chromosphere, except that the chromospheric power shows a strong correlation in the frequency range 3-3.5 mHz. 
\keywords{Sun: photosphere -- Sun: chromosphere -- Sun: Oscillations -- Sun: Magnetic fields -- \begin{large}                                                                                                \end{large}Sun: {\it Hinode}}
}

   \authorrunning{A. Raja Bayanna et al.}            
   \titlerunning{Relation between intensity oscillations and magnetic fields : Hinode observations}  

   \maketitle

%
%
\section{Introduction}           
\label{sect:intro}
Over the years, a wide range of oscillatory phenomena have been observed in various regions of the Sun. Oscillations in the solar atmosphere have been studied since 1960s (Leighton, Noyes \& Simon \cite{Leighton1962}). These studies have improved our understanding of the internal structure of the Sun as~well~as the dynamic structure of sunspots. Intensity, velocity and magnetic field observations of the Sun in various spectral lines have been used in studying these oscillations (Staude \cite{Staude1999}). 

The studies related to the photosphere emphasize on understanding the internal structure of the Sun through acoustic waves, while the studies with the chromospheric lines focus on understanding the propagation of waves to the higher atmosphere, their interaction with the magnetic fields there, and consequently understanding the problem of coronal heating.

The important findings of these studies in magnetic media are: (i) the presence of the photospheric five minute oscillations, and its absorption in the regions of strong photospheric magnetic fields (Braun~et~al. \cite{Braun1987}, \cite{Braun1988}, \cite{Braun1992}, \cite{Braun1993}; Bogdan~et~al. \cite{Bogdan1993}; Hindman \& Brown \cite{Hindman1998}; Kumar~et~al. \cite{Kumar2000}; Jain \& Haber \cite{Jain2002}; Venkatakrishnan~et~al. \cite{PVK2002}), (ii) enhanced oscillations in chromospheric umbra in the three minute band (Bhatnagar \& Tanaka \cite{Bhatnagar1972}; Lites \cite{Lites1986}; Kentischer \& Mattig \cite{Kenti1995}; Nagashima~et~al. \cite{Nagashima2007}), and (iii) running penumbral waves (Zirin \& Stein \cite{Zirin1972}; Giovanelli \cite{Giovanelli1972}, \cite{Giovanelli1974}; Maltby \cite{Maltby1975}; Christopoulou~et~al. \cite{Christopoulou1999}, \cite{Christopoulou2000}, \cite{Christopoulou2001}; Bloomfield~et~al. \cite{Bloom2007}). Some of the review articles (Bogdan \cite{Bogdan2000}, Solanki \cite{Solanki2003}, Bogdan \& Judge  \cite{Bogdan2006}) explain the work done on the nature of sunspot oscillations and related problems. In this context, simultaneous time-series observations in various spectral lines that sample the sunspot atmosphere at different heights using high resolution instruments such as the Solar Optical Telescope (SOT) (Tsuneta~et~al. \cite{HinodeSOT2008}) on board {\it Hinode} (Kosugi~et~al. \cite{Hinode2007}) and the Helioseismic and Magnetic Imager (Schou~et~al. \cite{HMI2012}) on board {\it Solar Dynamics Observatory} ({\it SDO}; Pesnell~et~al. \cite{SDO2011}) can be useful in studying these oscillatory processes and their contribution to the dynamics in the solar atmosphere. 
 
The high-resolution and multi-wavelength capability of {\it Hinode} provide several important opportunities to local helioseismologists. These allow to understand and confirm many physical processes in the sub-surface layers of the Sun and also in its atmosphere (Sekii \cite{sekii2009}; Kosovichev \cite{koso2012}), some of which are as follows. Nagashima~et~al. (\cite{Nagashima2007}) studied intensity oscillations in a sunspot and showed that G~band power is suppressed in sunspot umbra, while Ca~{\sc ii}~H observations revealed high-frequency oscillations with a peak at 6~mHz. Kosovichev \& Sekii (\cite{koso2007}) studied the flare-induced high-frequency chromospheric oscillations in a sunspot. Sekii~et~al. (\cite{sekii2007}) confirmed that the supergranulation is a shallow phenomenon. Similarly, Mitra-Kravev~et~al. (\cite{Mitra2008}) examined the phase difference between oscillations of the photosphere and chromosphere.  Zhao~et~al. (\cite{Zhao2010}) obtained travel-time measurements for short distances without phase-speed filtering and confirmed the sound-speed results, which were obtained using data from Michelson Doppler Imager (MDI) ( Scherrer~et~al. \cite{MDI1995}) on board the {\it Solar and Heliospheric Observatory} ({\it SoHO}; Domingo~et~al. \cite{Domingo95}) with the phase-speed filtering. 
 
We study intensity fluctuations in different regions of an active region and the correlation between the different parameters of the photospheric magnetic fields and the intensity oscillatory power at different heights in the solar atmosphere in different frequency bands using {\it Hinode}/SOT data. Earlier investigations carried out by Mathew (\cite{Mathew2008}) using Dopplergrams from {\it SoHO}/MDI and potential field computations from {\it SoHO}/MDI line-of-sight magnetograms revealed that the umbra-penumbra boundary showed enhanced absorption of power, where the transverse potential field was strongest. Gosain~et~al. (\cite{Gosain2011}) confirmed the aforementioned result by relating the power obtained from {\it SoHO}/MDI Dopplergrams with a vector magnetogram obtained from {\it Hinode}. Using high-resolution observations of G~band and Ca~{\sc ii}~H obtained from {\it Hinode}, Nagashima~et~al. (\cite{Nagashima2007}) studied the power in spatial scales corresponding to umbral flashes and they observed a node-like structure at the center of umbra with suppressed power in Ca~{\sc ii}~H power maps at all frequencies. To that end, we employ high temporal and spatial resolution observations from {\it Hinode} to study the nature of sunspot oscillations and its relation to the photospheric magnetic field parameters.

\section{The observational data}
\label{sec:obs}
 
\begin{figure}
\centering
\includegraphics[width=1.0\textwidth]{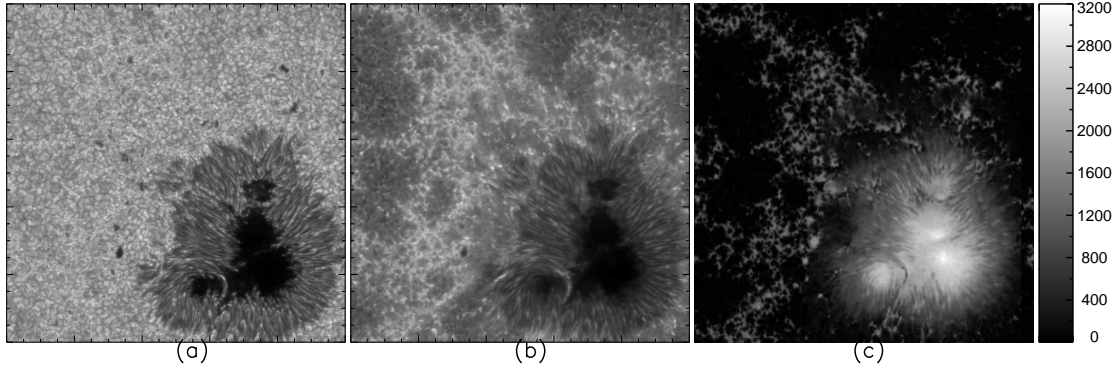}
\caption{Intensity filtergrams of the active region NOAA 10953 in: (a) G~band and (b) Ca II H line with a field-of-view of 112 arc-sec$^2$. The magnetic field strength map is shown in (c).}
\label{fig1}
\end{figure}

We have used a 3 hr 30 min sequence of G band and Ca II H filtergrams of the active region NOAA 10953 recorded by the Broad-band Filter Imager (BFI) of the {\it Hinode}/SOT to study the intensity oscillations in the active region. The filtergrams were acquired on 2007 May 1 during 14:31-17:57 UT and have a spatial sampling of $0\farcs11$ pixel$^{-1}$ and a cadence of one minute. The active region was located at S10W05 on the solar disk. The field-of-view (FOV) of the filtergrams is 112 arc-sec$^2$. G~band filtergrams were acquired nearly 3~s later to Ca~{\sc ii}~H filtergrams. In-addition to the broad band images, near simultaneous spectro-polarimetric observations of the active region from SOT/SP (Ichimoto~et~al. \cite{HinodeSP2008}) have been used in our analysis. SOT/SP records the four Stokes spectra of the Fe line pair at 630 nm with a spectral sampling of 21.5 m\AA~ and an exposure time of 4.8 s at each slit position. We have used level-2 maps comprising magnetic field strength, inclination, and azimuth which were obtained by inverting the observed Stokes profiles employing the MERLIN\footnote{Level-2 data of the active region was made available by the Community Spectro-polarimetric Analysis Center(CSAC) at HAO.} code. The active region was scanned in the fast mode with a step width of 0\farcs29 and a sampling of 0\farcs32 along the slit. These maps were interpolated to a spatial scale of 0\farcs32~pixel$^{-1}$ in both the directions. Consequently, the images obtained with BFI were also re-scaled to 0\farcs32~pixel$^{-1}$ resolution to match that of the SOT/SP maps.
 
 \section{Analysis and Results}
 
The images were corrected for flat-fielding, dark current and bad pixels using standard Solarsoft routines. Although, the correlation tracker is employed to take care of global motion of the region of interest, we performed a rigid alignment of the sunspot as a function of time. This was done using a FFT based 2D cross-correlation routine, updating the reference frame at every 10$^{\textrm{\tiny{th}}}$ frame to account for the evolution of the sunspot. Figure~\ref{fig1} shows the snapshot of the active region in G~band and Ca~{\sc ii}~H along the map of magnetic field strength. Figure~2 shows time averaged G~band and Ca~{\sc ii}~H images deduced over the period 14:31-17:57 UT  on 2007 May 1. The average images were normalized with their exposure times to obtain the counts in same scale in both the images.

A two point backward difference filter (Garc{\'{\i}}a \& {Ballot}, \cite{Garcia2PBF}) was applied to obtain the first difference of the time series and the filtered data were normalized by the mean intensity in the two running frames as shown in the equation (1). First difference enhances the oscillatory signals above the background variations and the normalization by the mean intensity causes a smooth transition between the umbra and the penumbra (Nagashima~et~al. \cite{Nagashima2007}).
 
\begin{eqnarray}
\hat{I_k} = 2\left(I_k-I_{k-1}\right)/\left(I_k+I_{k-1}\right)
\end{eqnarray}
Where, $\hat{I_k}$ and $I_k$ are normalized intensity and intensity of the $k^{th}$ image of the sequence, respectively.

Further, we computed the Power Spectral Density (PSD) from the mean normalized differential intensity fluctuations at each pixel and generated 3D power maps with frequency along the z-direction. PSD in each pixel is corrected for $\omega^2$ to remove the effect of the time-derivative (Nagashima~et~al. \cite{Nagashima2007}). The variation of intensity oscillatory power in different regions of the active region in G~band and Ca~{\sc ii}~H observations are studied in the following sections.

\begin{figure}[b]
\centering
\includegraphics[width=0.85\textwidth]{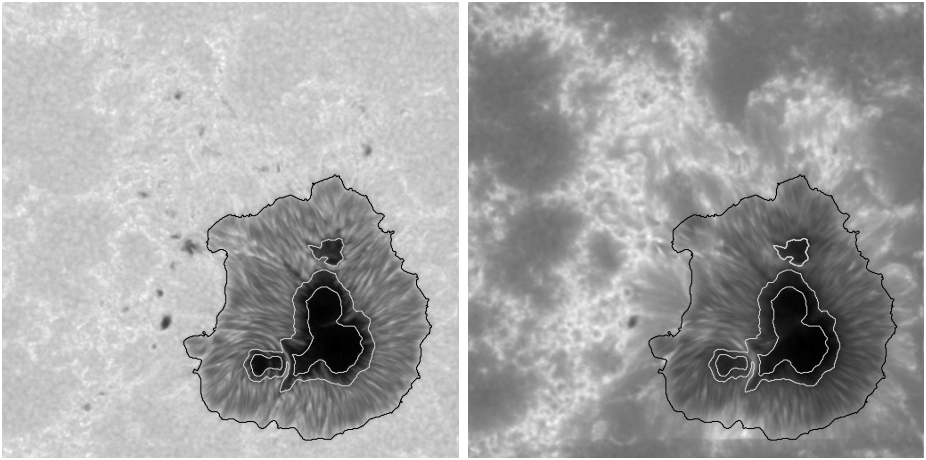}
\caption{Average intensity maps of the active region in G~band (left) and Ca~{\sc ii}~H line (right). The contours correspond to umbra, umbra-penumbra boundary, penumbra and quiet Sun. The bright regions outside the sunspot are considered as plage while the regions that are neither bright in G band and Ca~{\sc ii}~H are assumed to be quiet Sun.} 

\label{fig3}
\end{figure}
 
\subsection{Power distribution in different regions of the active region and quiet Sun}
\label{sec:3}
 
\begin{figure}[t]
\centering
\includegraphics[width=0.425\textwidth]{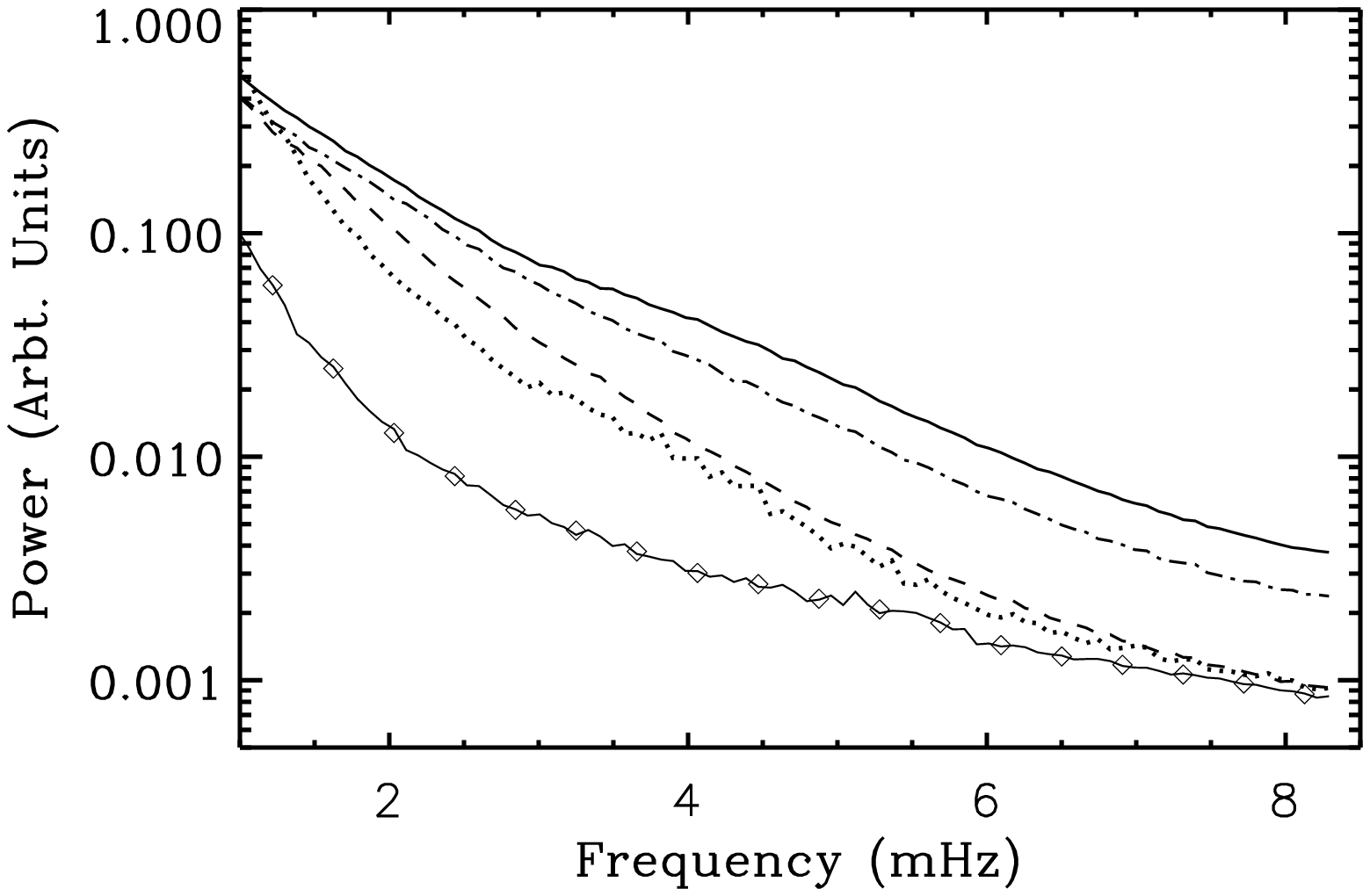}
\includegraphics[width=0.425\textwidth]{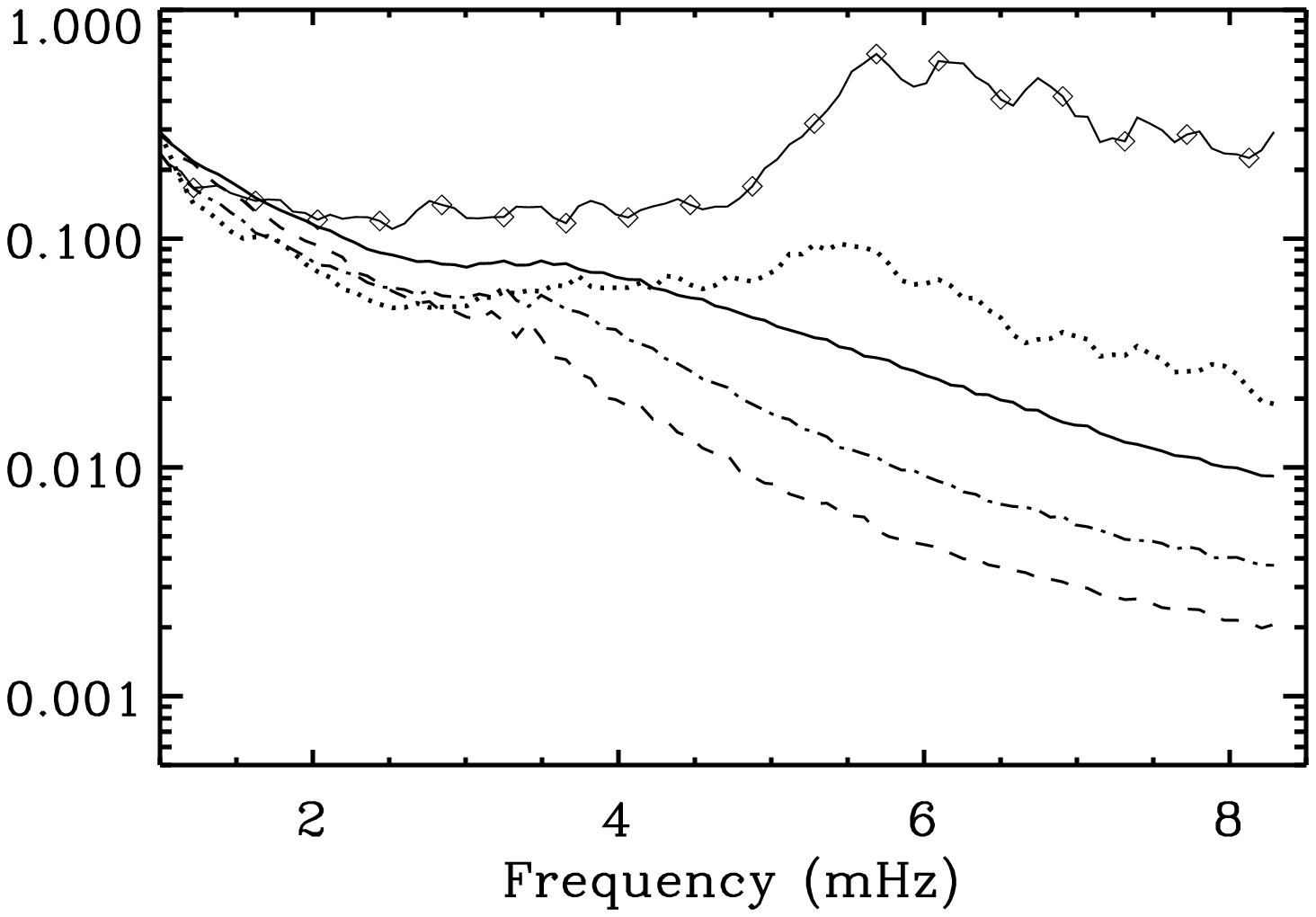}
\caption{Power distribution of G~band intensity (left) and Ca~{\sc ii}~H intensity (right) in different regions of the FOV. Power in umbra (solid line with diamond symbols), umbra-penumbra boundary (dotted line), penumbra (dashed line), plage (dot-dashed) and quiet Sun (solid line) are shown here.} 
\label{layerwise}
\end{figure}

\begin{figure}[b]
\centering
\includegraphics[width=0.275\textwidth]{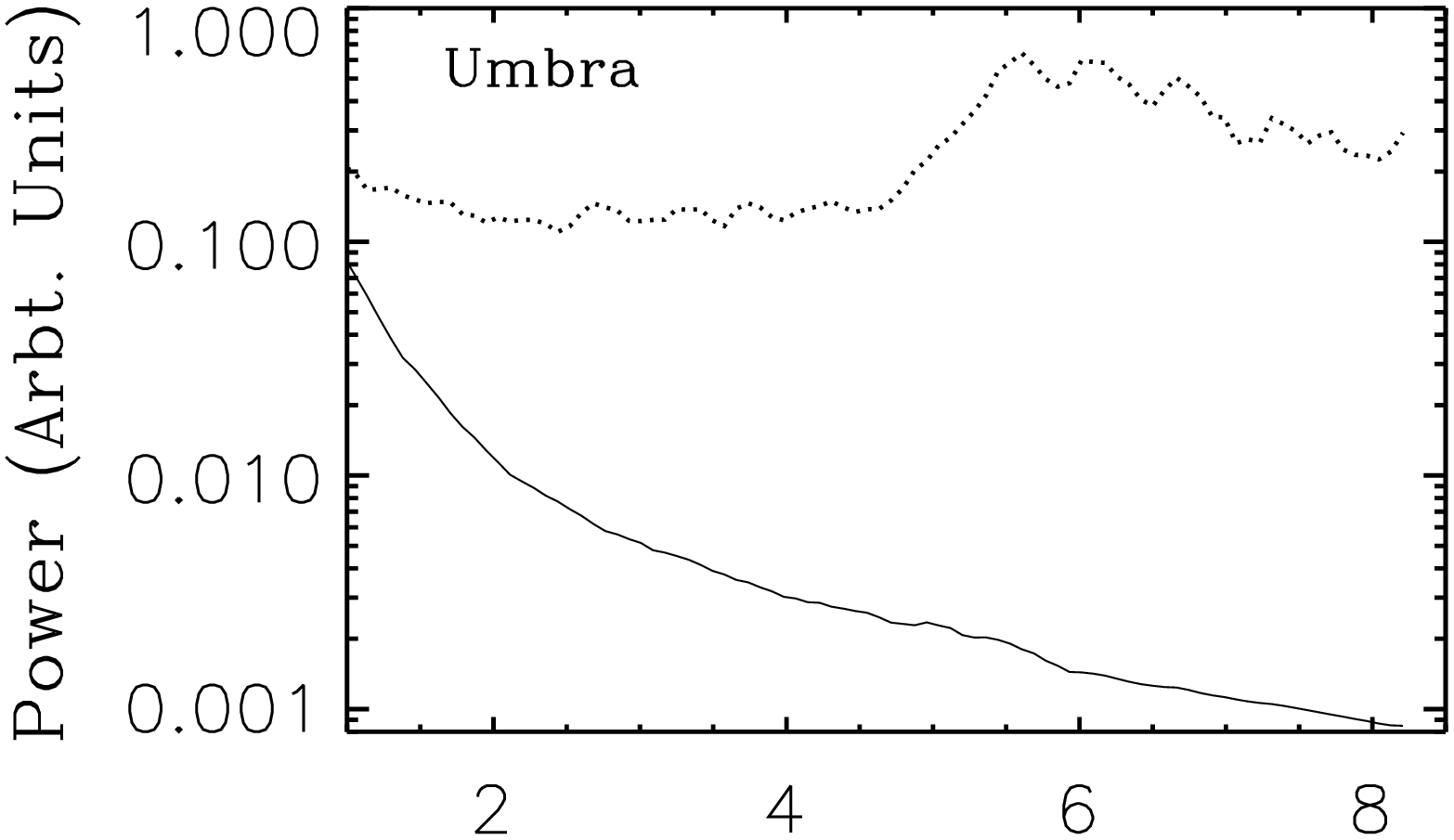}
\hspace{1mm}
\includegraphics[width=0.275\textwidth]{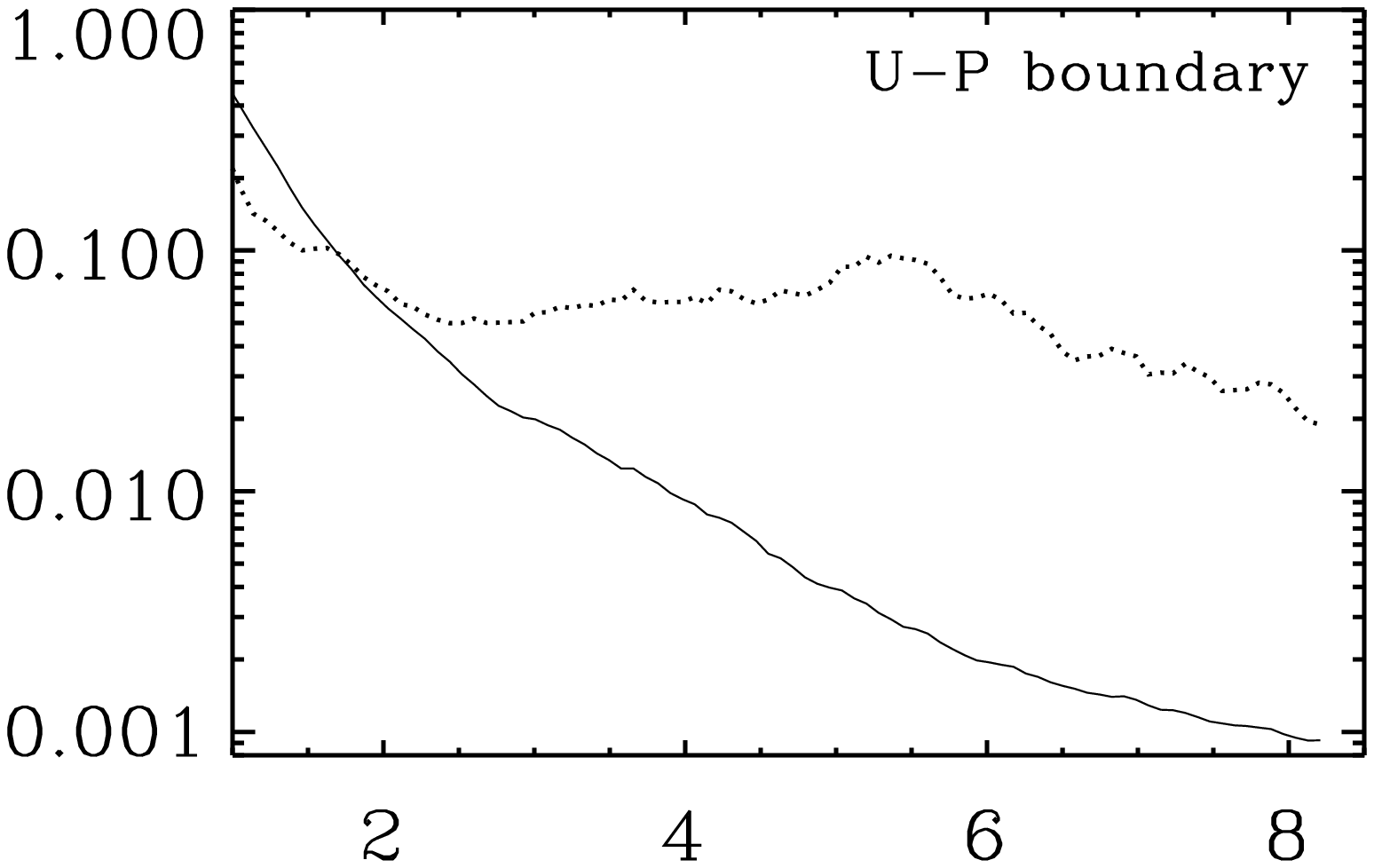}
\hspace{0.5mm}
\includegraphics[width=0.275\textwidth]{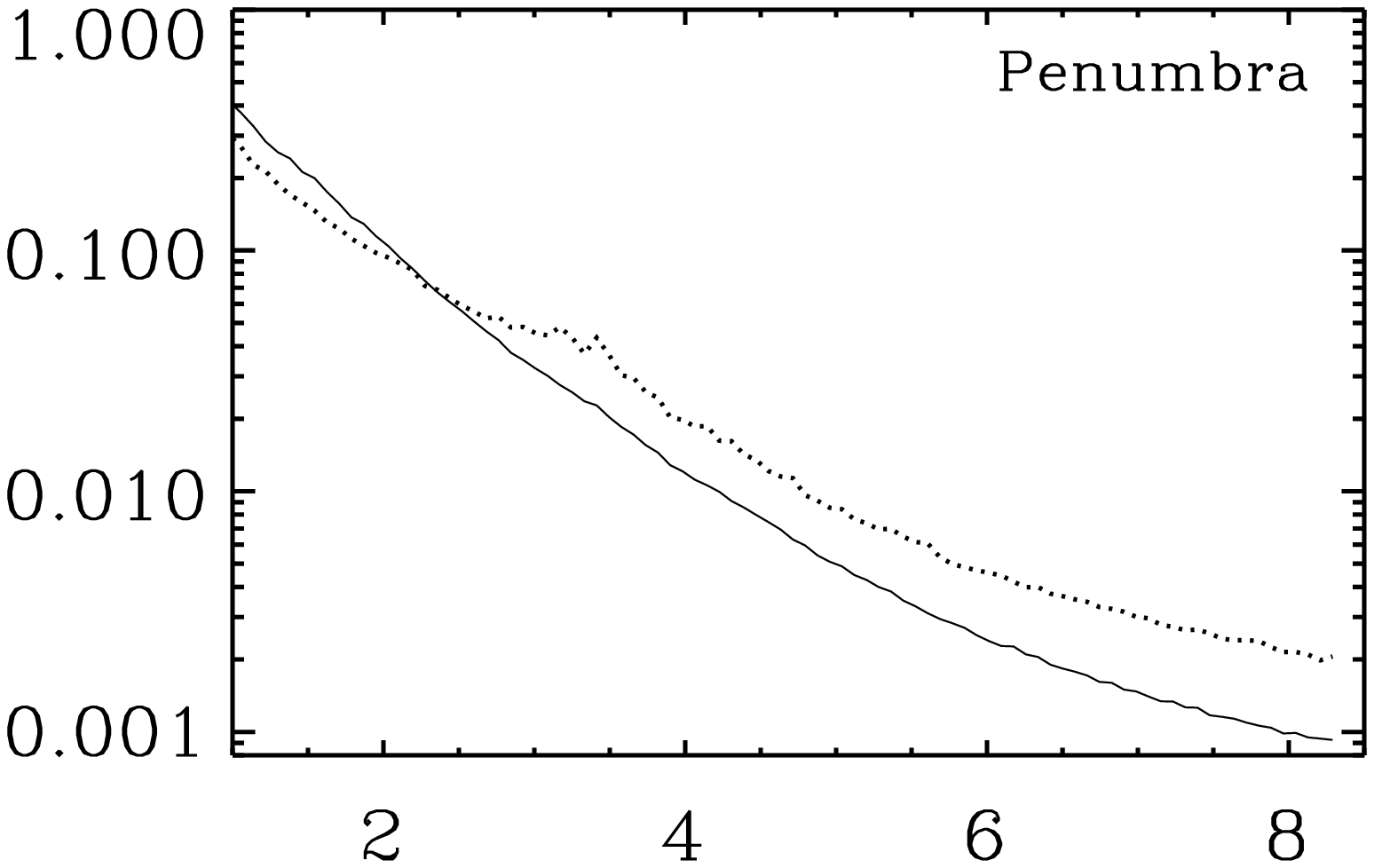}
\includegraphics[width=0.275\textwidth]{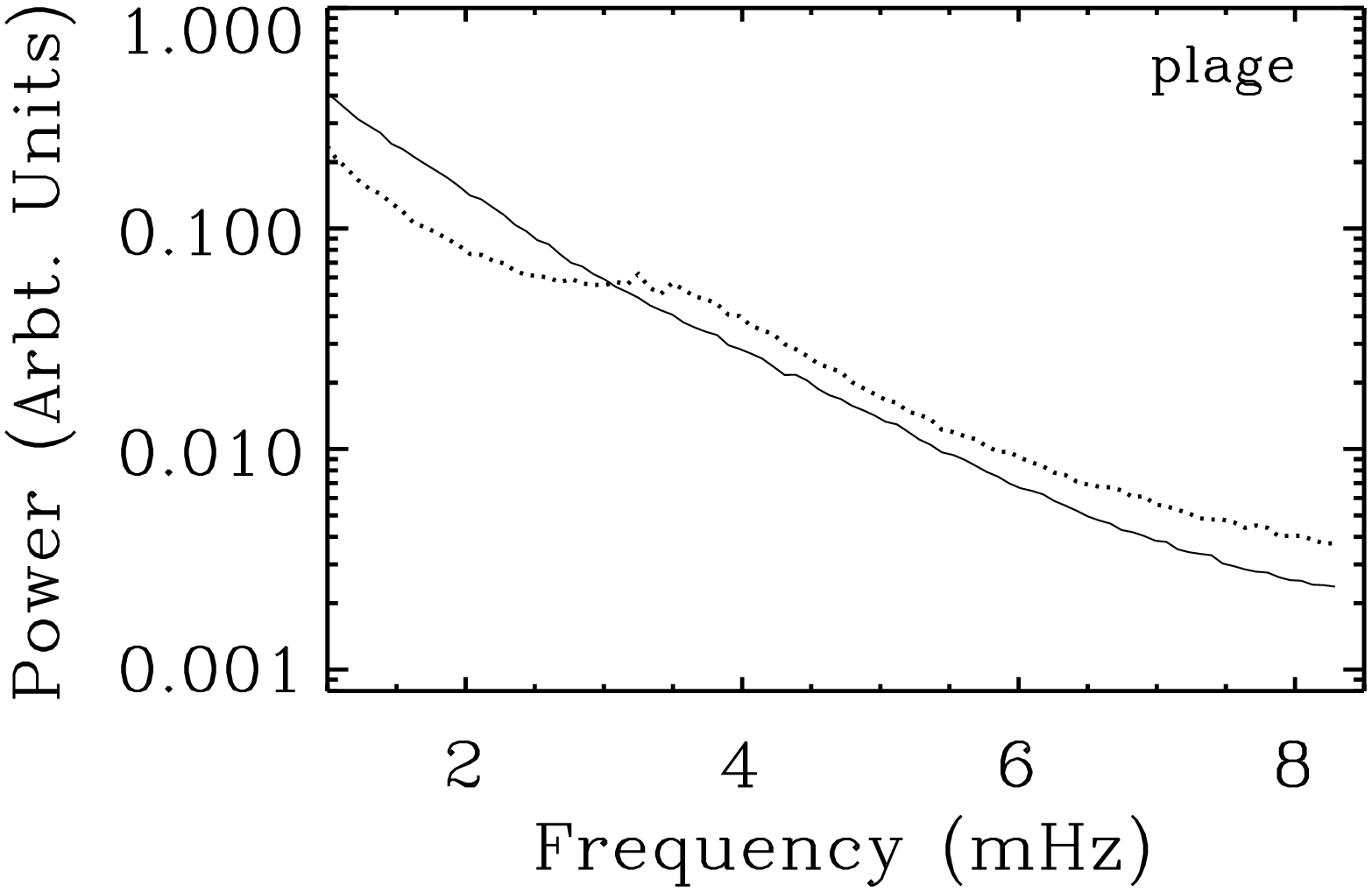}
\hspace{0.5mm}
\includegraphics[width=0.275\textwidth]{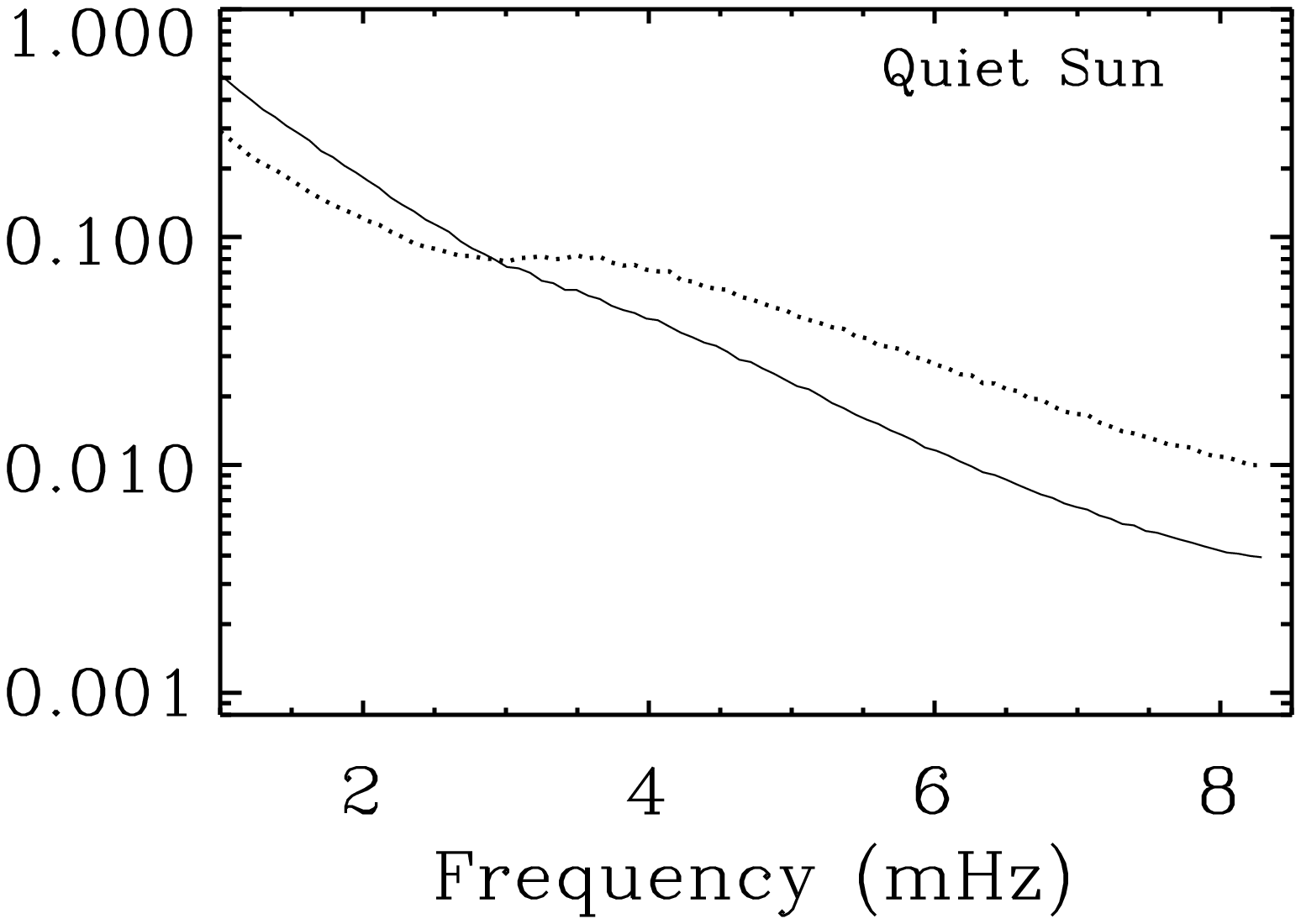}
\hspace{1mm}
\includegraphics[width=0.275\textwidth]{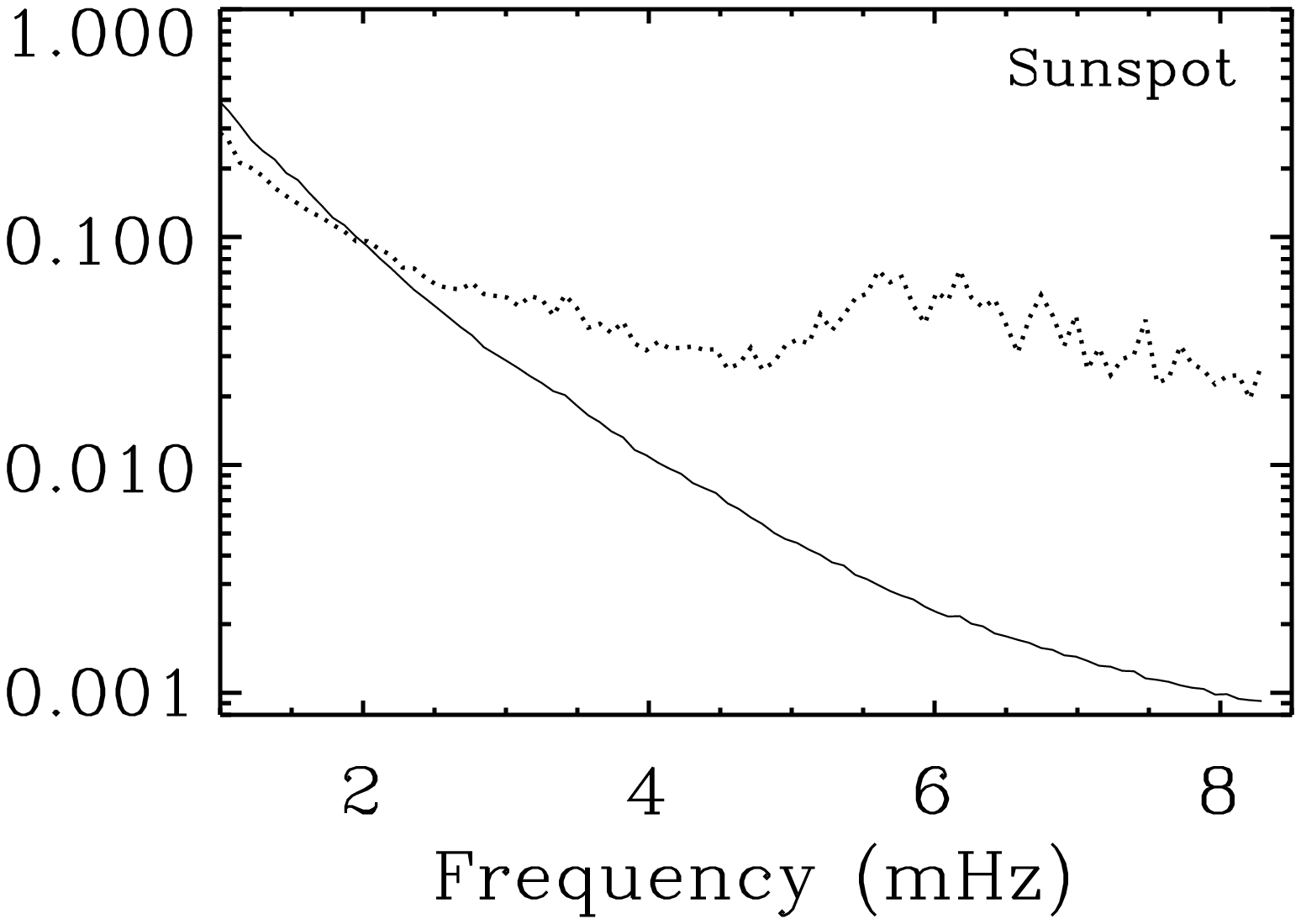}
\caption{Power distribution of G~band intensity (solid line) and Ca~{\sc ii}~H intensity (dotted line) in the regions: umbra (top-left), umbra-penumbra boundary (top-middle), penumbra (top-right), plage (bottom-left), quiet Sun (bottom-middle) and sunspot (bottom-right).} 
\label{Rwise}
\end{figure}

In order to investigate the power distribution in different regions of the active region, the FOV was divided into the following regions: umbra, umbra-penumbra boundary (UPB), penumbra, plage, and quiet Sun. The iso-intensity contours overlaid on the time averaged G-band and Ca~{\sc ii}~H images in Figure 2 were determined from the peaks in the intensity distribution. The contours enclosing the umbra, UPB, and penumbra were obtained from the time averaged G-band image while the same for the plage region was determined from the time averaged Ca image; the bright regions outside the sunspot are considered as plage while the regions that are neither bright in G band and Ca~{\sc ii}~H are assumed to be quiet Sun. The fraction of pixels corresponding to umbra, UPB, and penumbra are 0.022, 0.030, 0.22, respectively. Plage region and quiet region contain 0.626 and 0.091 fraction of pixels, respectively. The average power in each of these regions as a function of frequency is shown in Figures~\ref{layerwise}~and~\ref{Rwise}. Figure~\ref{layerwise} allows us to compare the power in the different regions described above, separately in G~band and Ca~{\sc ii}~H. On the other hand, Figure~\ref{Rwise} shows a comparison between the powers in G~band and Ca~{\sc ii}~H in each of the regions.

It is observed that intensity oscillatory power in G~band decreases from the quiet Sun to umbra (i.e., in the order quiet Sun, plage, penumbra, UPB, and umbra) at all frequencies. On the other hand, power in Ca~{\sc ii}~H shows such a trend from quiet Sun to penumbra only, with the exception of slight enhancement of power in 3-4 mHz frequency range (c.f., Figure ~\ref{layerwise}). It is also observed that overall power is lower in G~band as compared to Ca~{\sc ii}~H with a cross-over seen at 3~mHz which shifts to 0~mHz as we move from the quiet Sun to the umbra of the sunspot (c.f. Figure~\ref{Rwise}). Thus, Ca~{\sc ii}~H oscillations are seen to be richer in high-frequency power in the magnetized environment, which is in good agreement with the results of Nagashima~et~al. (\cite{Nagashima2007}). We also observe the presence of 5-minute oscillations in  quiet Sun, plage and penumbral regions.
 
\subsection{Intensity power maps as function of magnetic field strength and frequency}

\begin{figure}[h]
\centering
\includegraphics[width=0.40\textwidth]{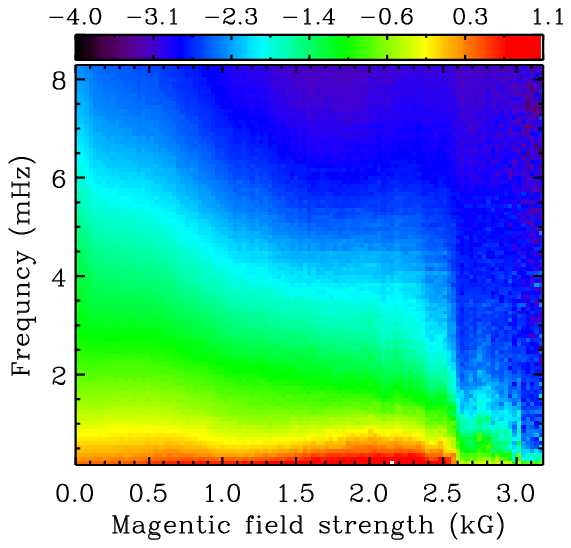}
\hspace{5mm}
\includegraphics[width=0.40\textwidth]{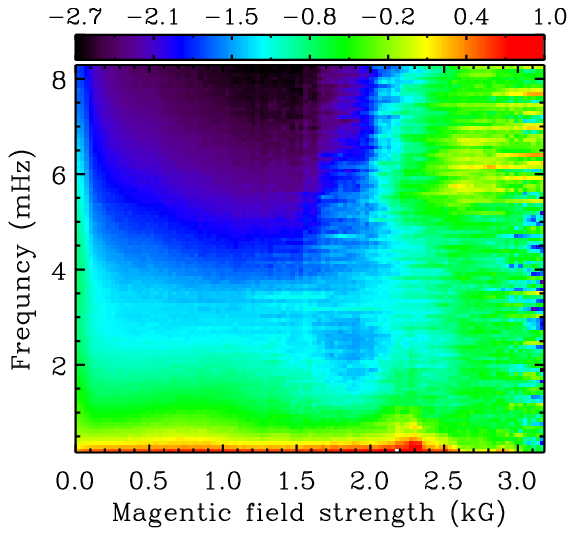}
\caption{Power maps from the intensity variations in the active region NOAA 10953 in G~band (left) and Ca~{\sc ii}~H (right) as function of magnetic field strength and frequency. The maps are shown in logarithmic scale as indicated by the color bar.}
\label{magpow}
\end{figure}

In order to examine the relation between oscillatory power in G~band and Ca~{\sc ii}~H with respect to the photospheric magnetic field strength ($B$) and frequency, we constructed power maps as function of magnetic field strength and frequency as indicated in Figure~\ref{magpow}. These maps were derived by averaging the power in all pixels having similar magnetic field strengths over a 30~G interval at each frequency ranging from 0.1 to 8.3~mHz. We observe the following from these power maps.

\begin{enumerate}
\item In general, Ca~{\sc ii}~H shows larger power in comparison to that of G~band for any magnetic field strength and frequencies above 1~mHz. 
\item Power in G~band decreases with increase in magnetic field strength at all frequencies. 
\item Ca~{\sc ii}~H shows larger power in stronger magnetic fields ($|B|$~\textgreater 2200~G) at all frequencies and less power in the intermediate field regime (100~G~\textless~$|B|$~\textless~2000 G) at frequencies above 5~mHz. Weaker fields ($|B|$~\textless~100~G) show larger power in the frequency range 0-6~mHz in comparison to the same in the intermediate field strength regime.
\item Power map of Ca~{\sc ii}~H is more structured than that of G~band. It shows the signature of 5-minute oscillations for the magnetic field strengths between 1200-2000 G. These fields are mostly located in the penumbral region where the inclination is in the range of 90-130 degrees. In the higher magnetic field regions (mostly, umbra) it also shows oscillatory behavior.
\end{enumerate}

\begin{figure}[t]
\centering
\includegraphics[width=0.575\textwidth]{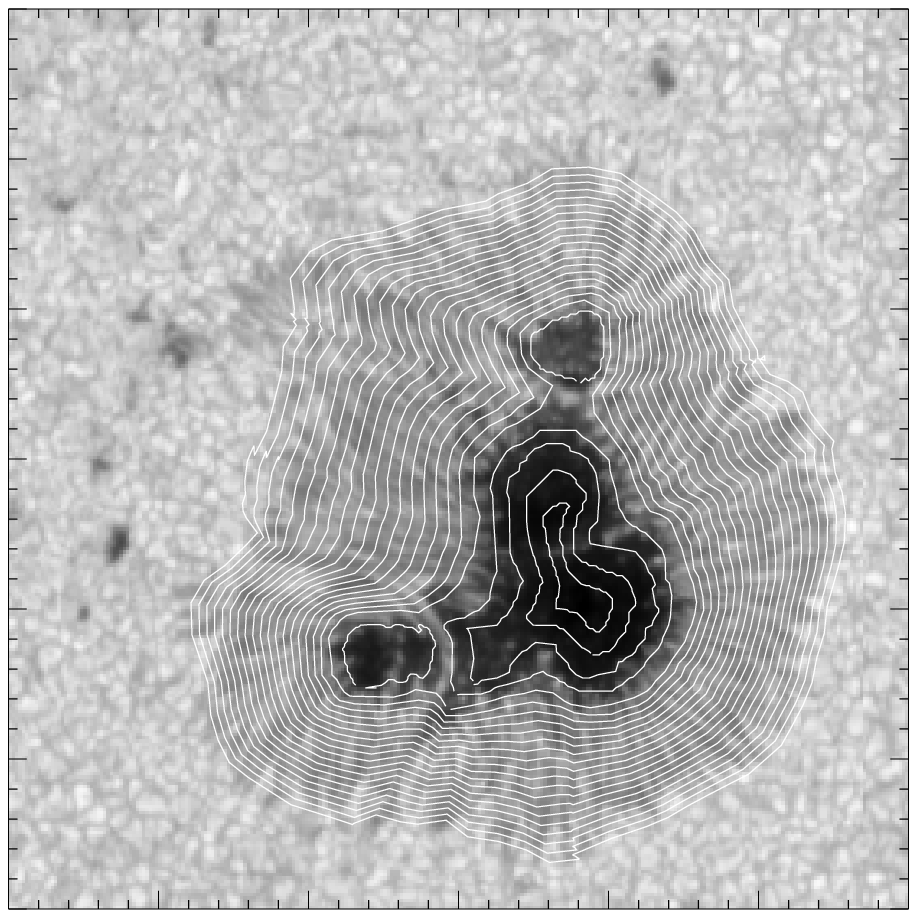}
\includegraphics[width=0.375\textwidth]{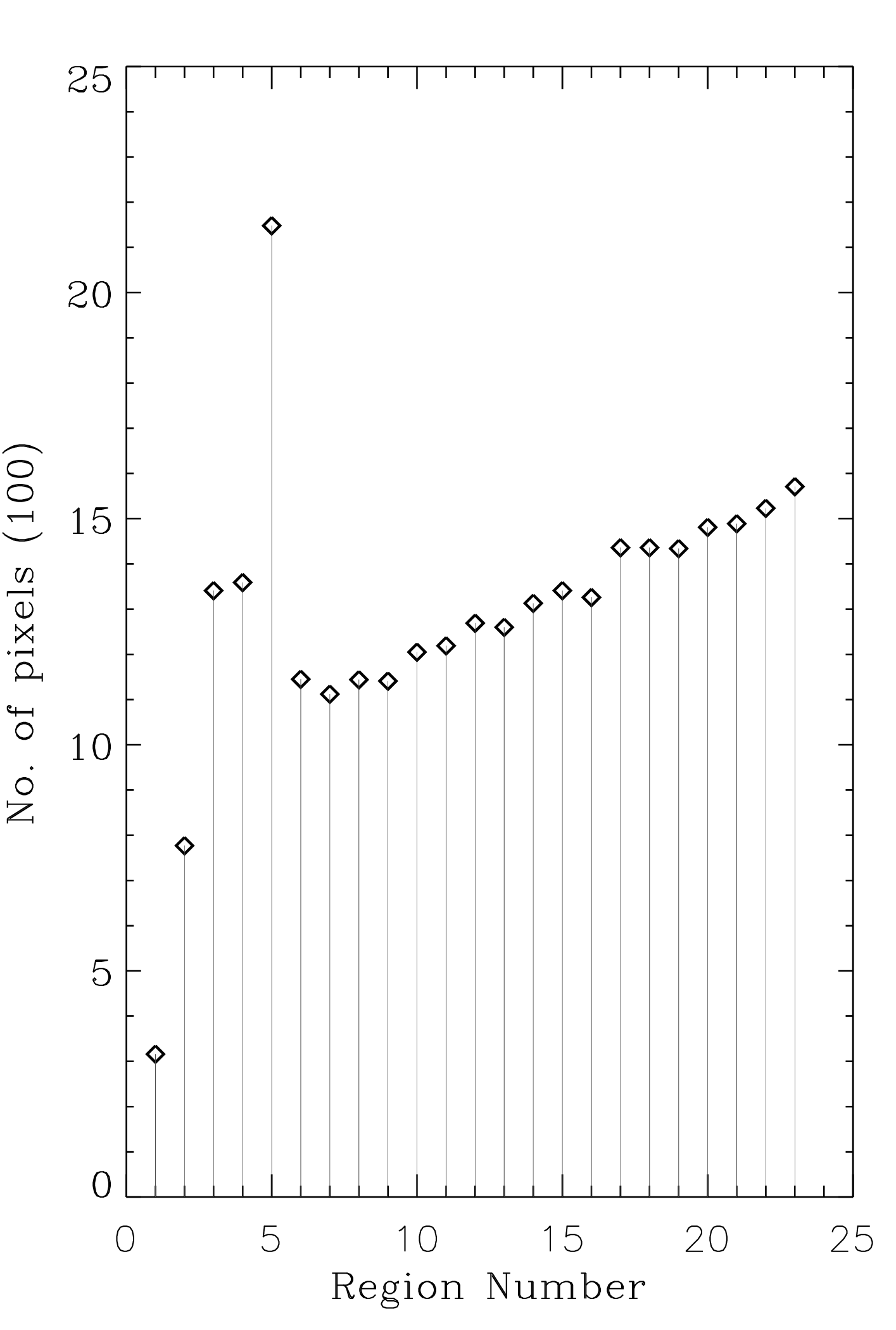}
\caption{Left: Enclosed curves overlaid on the intensity filtergram of the active region to study the radial and azimuthal variation of the magnetic field parameters and oscillations in various regions of the sunspot. Right: The number of pixels enclosed by each annular-like region are shown in the plot. The regions are numbered from center of umbra to sunspot boundary.}
\label{contou}
\end{figure}

\subsection{Relation between magnetic field parameters and oscillatory power in the sunspot}
\label{sec3}

The radial variation of the observed magnetic field parameters is studied using the enclosed curves shown in Figure~\ref{contou}. The umbra-penumbra boundary and penumbra-quiet Sun boundary obtained from the time-averaged G band image (c.f., Fig. 2), were used to construct 19 equidistant annular-like regions in the azimuthal direction which yielded the curves shown in Figure~\ref{contou}. The curves inside the umbra were hand drawn, which are separated by $\approx$1$^{\prime\prime}$. The other curves between umbra-penumbra boundary and penumbra are spaced apart by 0\farcs6. The regions 1-3 and 8-20 correspond to umbra and penumbra, respectively, while the regions 4-7 represent the umbra-penumbra boundary. The number of pixels enclosed by each region is illustrated in Figure~\ref{contou}.

\begin{figure}[t]
\centering
\includegraphics[width=0.30\textwidth]{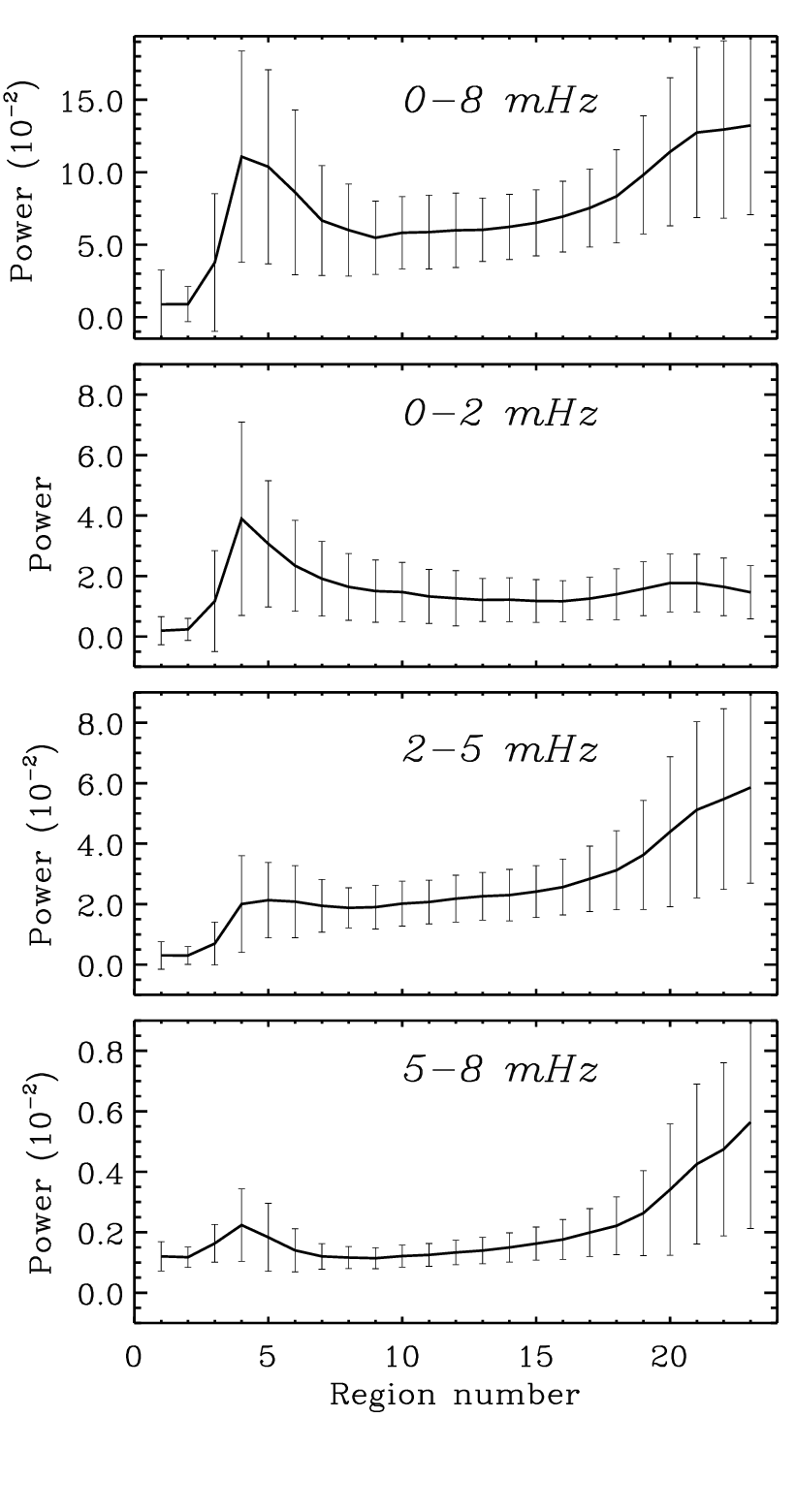}
\hspace{1mm}
\includegraphics[width=0.30\textwidth]{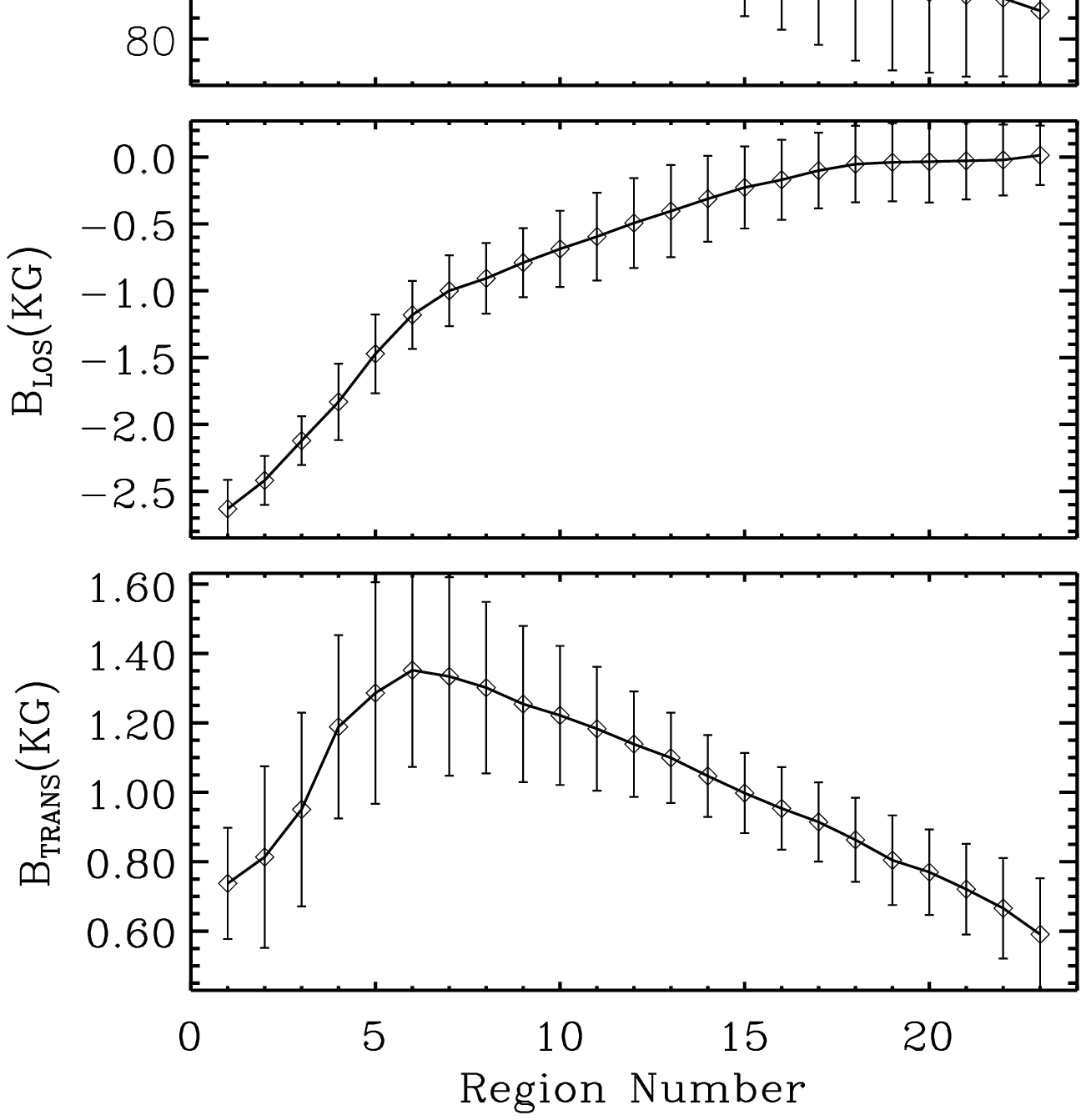}
\hspace{1mm}
\includegraphics[width=0.30\textwidth]{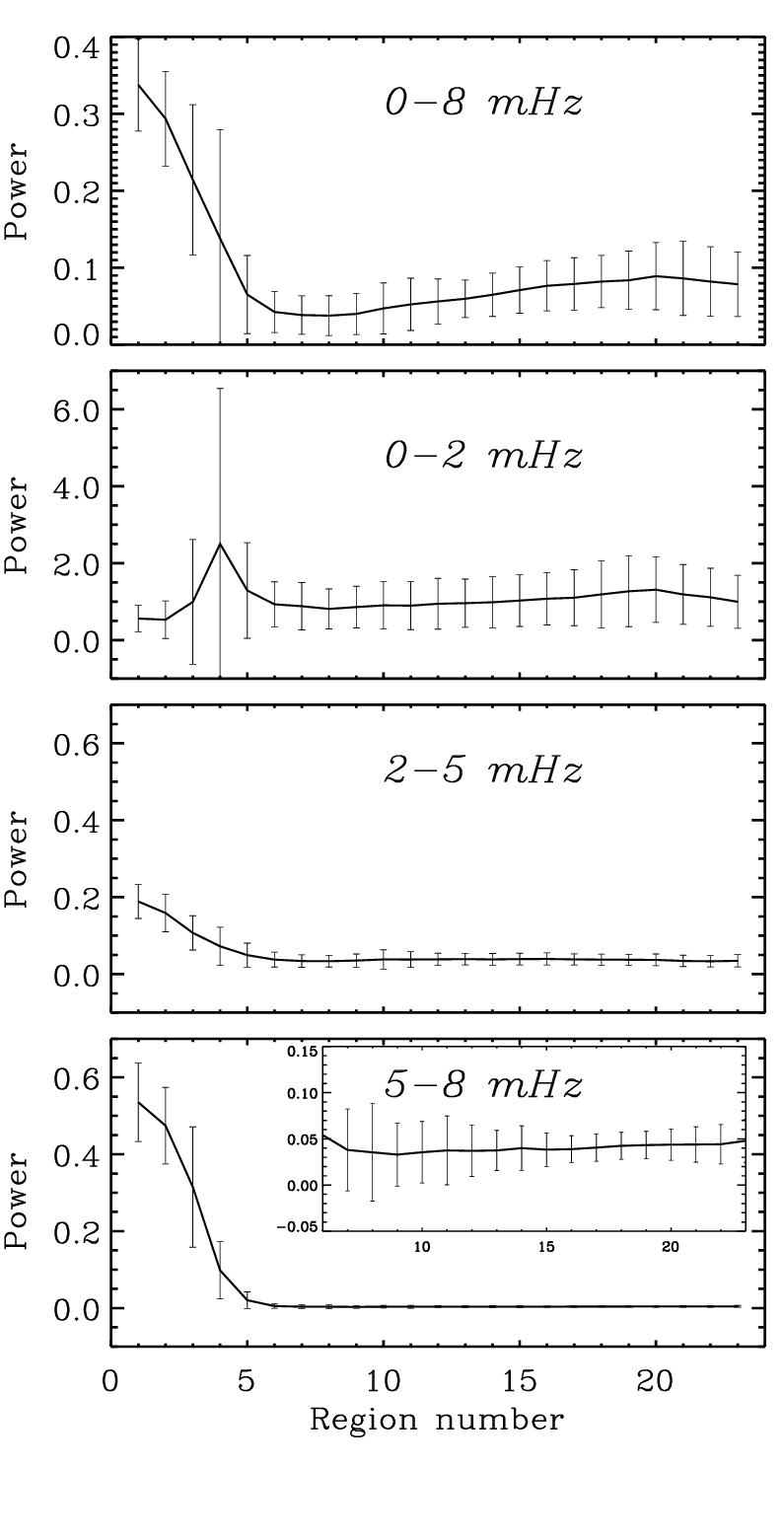}
\caption{Radial variation of power in G~band and Ca {\sc ii}~H are shown in left and right panels, respectively. Corresponding variation of magnetic field parameters ($B$, $\gamma$, $B_l$ and $B_t$) are shown in the middle panels. The power is averaged over frequency regimes: 0-8~mHz, 0-2~mHz, 2-5~mHz and 5-8~mHz. The vertical bars show $\pm1\sigma$ errors in the estimated values. The regions 1-3, 4-7 and 8-20 correspond to umbra, umbra-penumbra boundary and penumbra, respectively. The regions 21-23 correspond to sunspot-quiet Sun boundary.} 
\label{SPvar}
\end{figure}

\begin{figure}[t]
\centering
\includegraphics[width=0.425\textwidth]{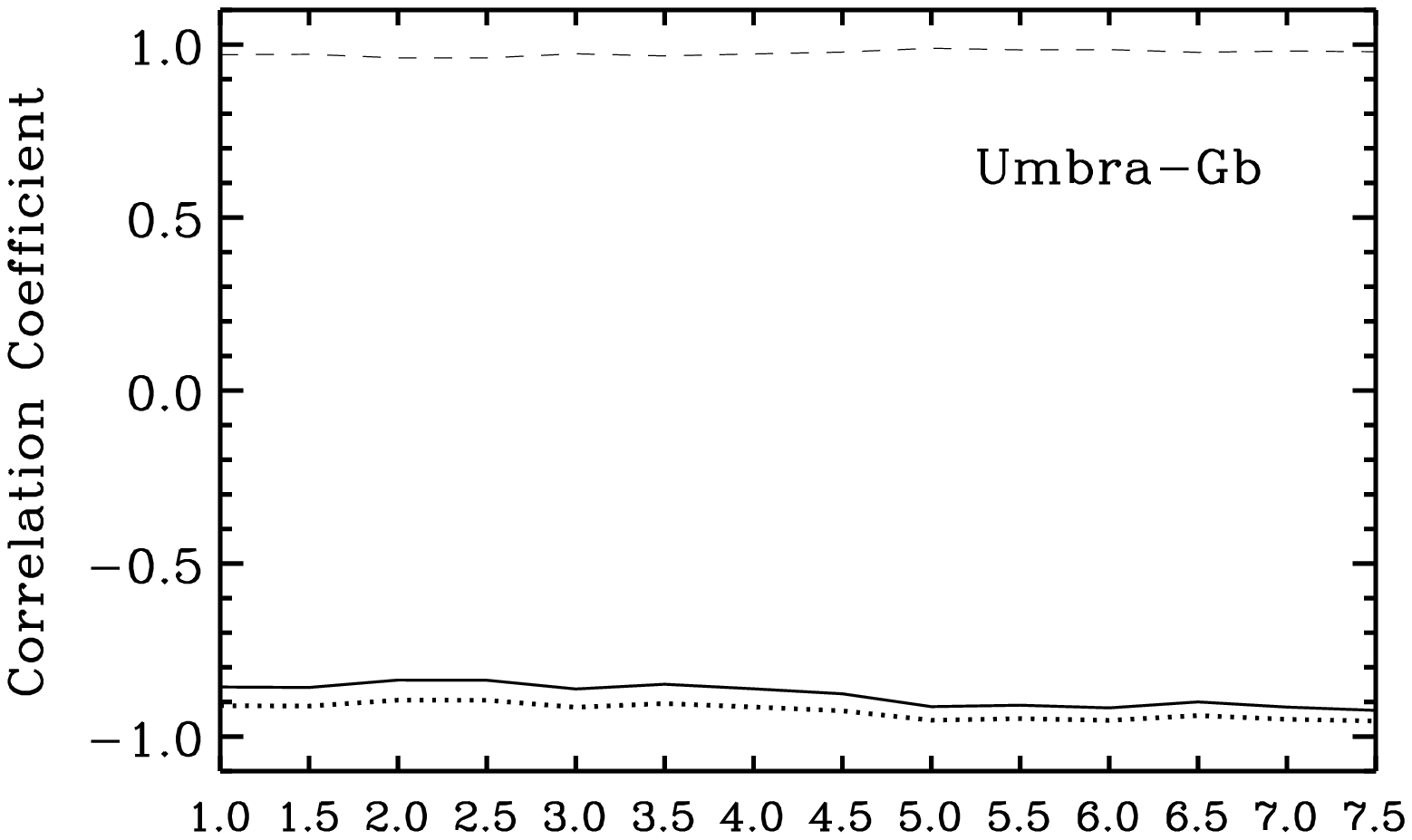}
\includegraphics[width=0.425\textwidth]{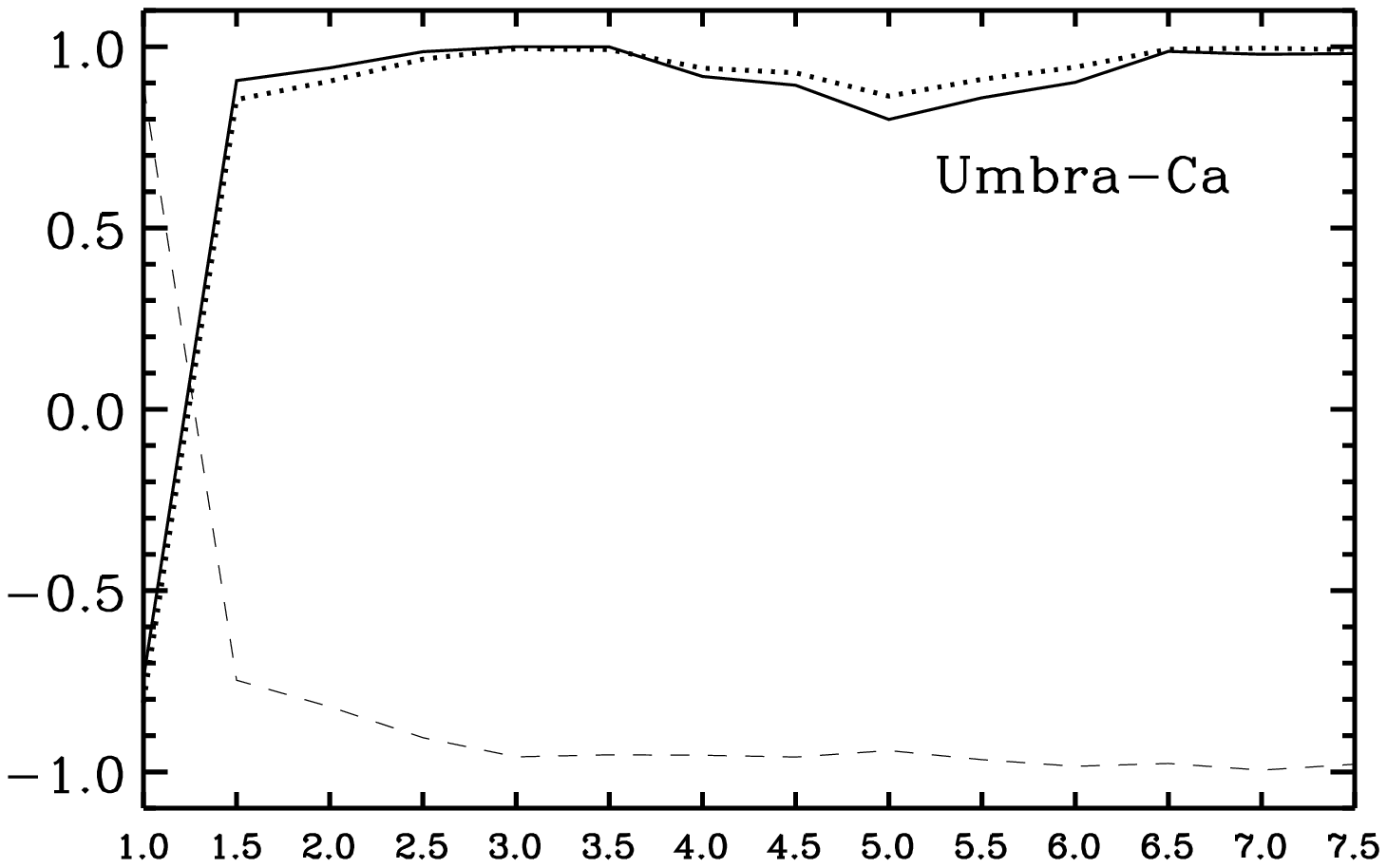}
\includegraphics[width=0.425\textwidth]{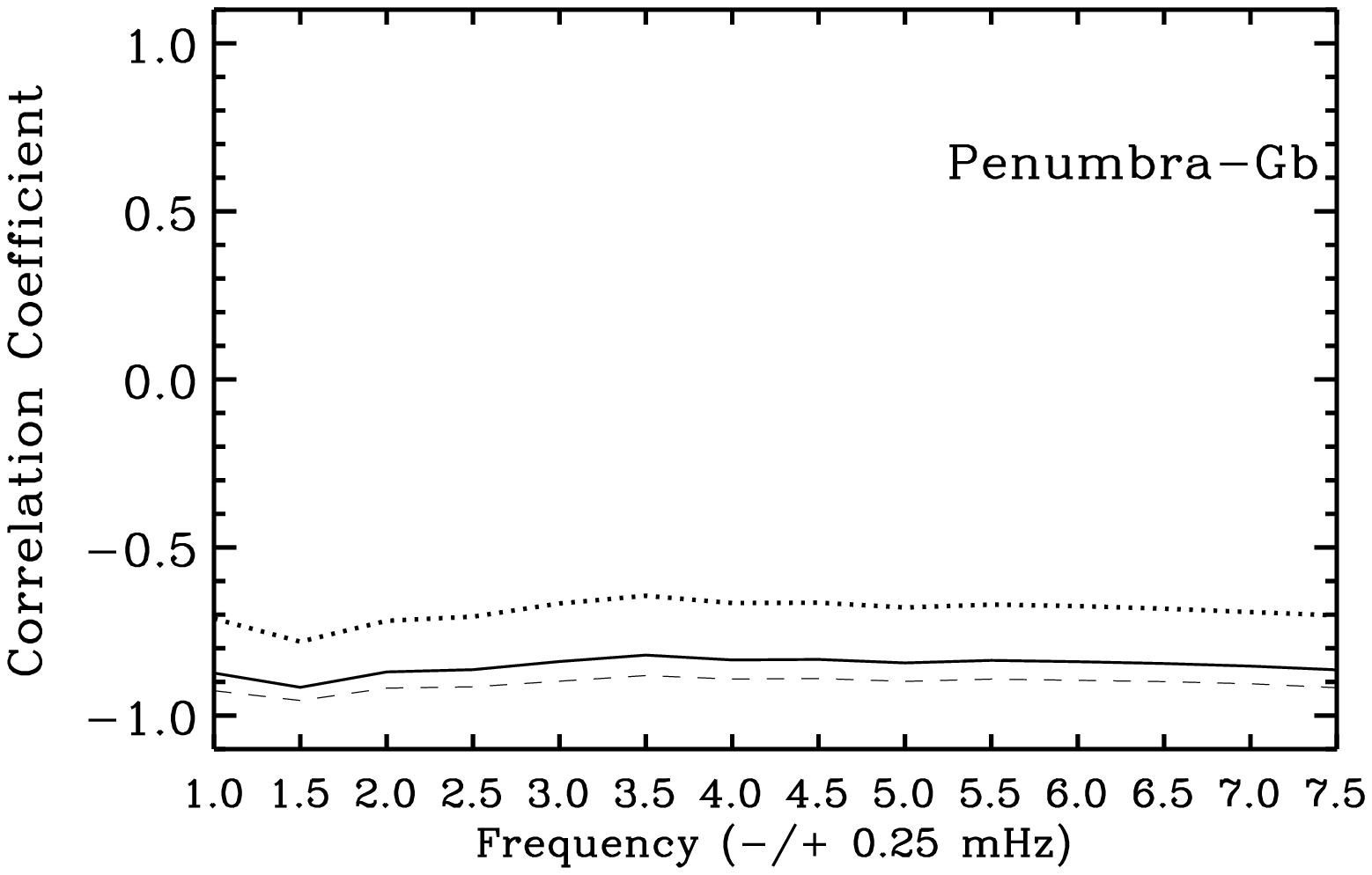}
\includegraphics[width=0.425\textwidth]{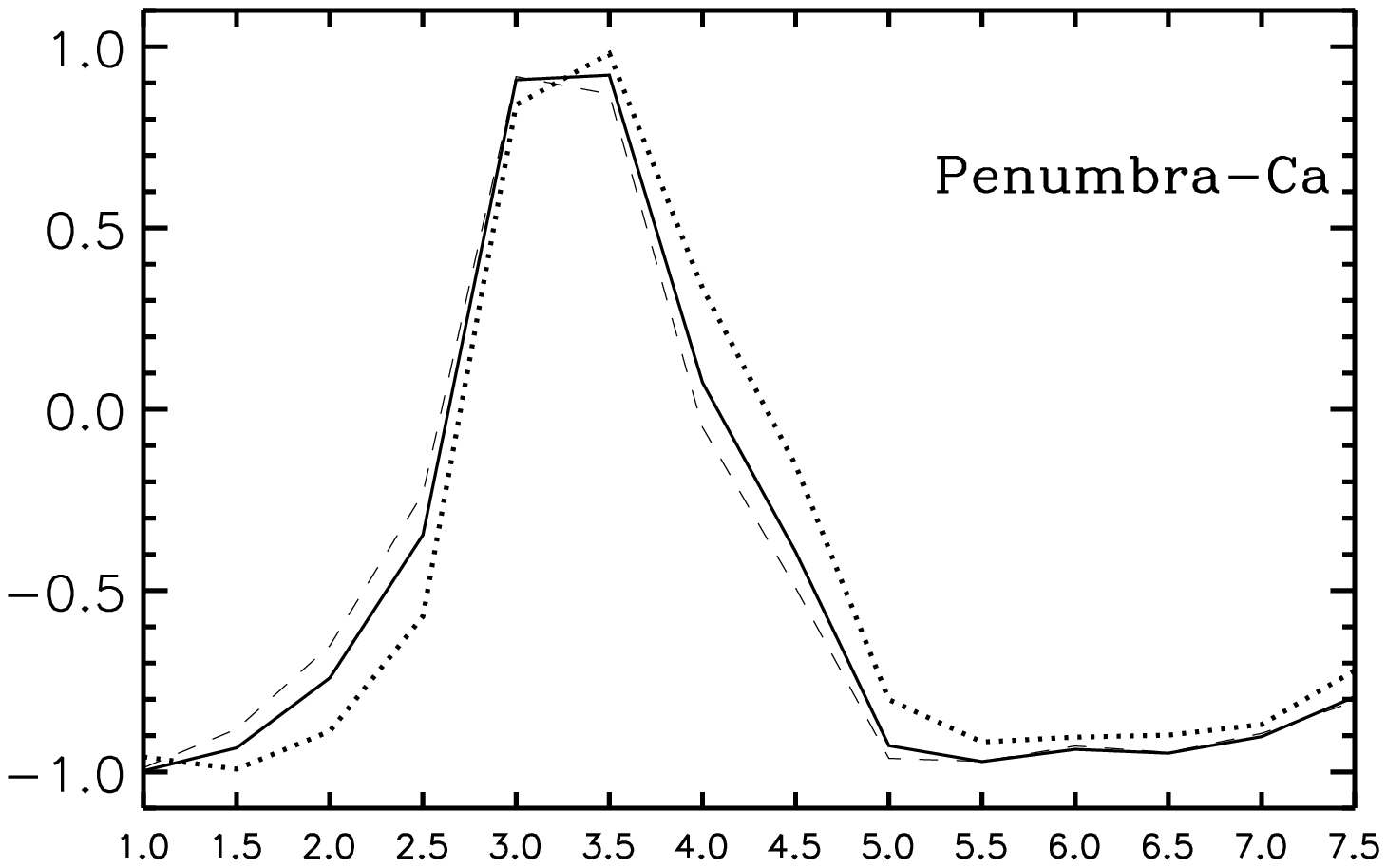}
\caption{Plots show correlation between the photospheric magnetic field parameters and intensity oscillatory power of umbra (top-panels) and penumbra (bottom panels) in G~band (left panels) and Ca~{\sc ii}~H (right panels) in different frequency bands. Correlation between $B$ vs. power (solid line), $B_t$ vs. power (dashed line), and $B_l$ vs. power (dotted) are shown here.} 
\label{Corr}
\end{figure} 

The azimuthally averaged radial profiles of various magnetic field parameters are shown in Figure~\ref{SPvar}. It is observed that magnetic field strength ($B$), inclination ($\gamma$), and line-of-sight ($B_l$) components of magnetic field show smooth variation with the radial distance. Transverse component of magnetic field ($B_t$) increases from center of the umbra to the inner-mid penumbra and, thereafter, this trend reverses. The vertical bars correspond to $\pm1\sigma$ errors in the estimated values. In general, the power in umbra-penumbra boundary (regions 4-7) and sunspot-quiet Sun boundary (regions 21-23) shows relatively larger spread in G~band. Whereas in Ca~{\sc ii}~H, the power in umbra and umbra-penumbra boundary shows larger spread.

The radial variation of power in Ca~{\sc ii}~H is dominant in the umbra in comparison to the other regions of the sunspot in the frequency regimes 2-5~mHz and 5-8~mHz. On the other hand, in 0-2~mHz band umbra-penumbra boundary shows enhancement of power both in Ca~{\sc ii}~H and G~band. Umbra-penumbra boundary in Ca~{\sc ii}~H shows reduction of power in 2-5~mHz and 5-8~mHz bands, while it shows enhancement of power in G~band for the above frequency ranges. This reduction/enhancement of power at umbra-penumbra boundary occurs around the peak value of the transverse magnetic field. The inclination of magnetic field at this location is about $130^\circ$. On the other hand, power estimated from the photospheric Dopplergrams obtained by MDI/{\it SoHO} showed enhanced {\it p-}mode absorption near umbra-penumbra boundary, where the inclination angle is nearly $135^\circ$ (Mathew~\cite{Mathew2008}; Gosain~et~al.~\cite{Gosain2011}).
 
In order to understand the relation between magnetic field parameters and intensity oscillatory power of the sunspot, we estimated the correlation-coefficients between radial variation of power and the magnetic field parameters averaged over each region. Here, it is observed that none of the magnetic field parameters ($B$, $B_l$, and $B_t$) show any correlation with the power in the sunspot. A Similar analysis by Gosain~et~al.~(\cite{Gosain2011}) did not reveal any strong association of the magnetic field parameters with Doppler power in a sunspot. However, when we derive the correlation-coefficients separately for the umbra and penumbra, we find distinct relations between the magnetic field parameters and the PSD which are shown in Figure~\ref{Corr}. When we consider sunspot as a whole, the distinct behaviors in the umbra and penumbra gets mixed-up and thus there is no correlation seen in our analysis.  

Intensity oscillatory power in Ca~{\sc ii}~H penumbra shows anti-correlation with $B$, $|B_l|$ and $B_t$ at all the frequencies except in the frequency range of 3.0-3.5~mHz, i.e. in the 5-minute regime. G~band intensity power in penumbra also shows anti-correlation with magnetic field  parameters. Similarly, the umbral power in Ca~{\sc ii}~H shows correlation with $B$, and $|B_l|$ above 1.5~mHz and opposite relation with $B_t$. Whereas, G~band umbral power shows completely opposite trend; i.e. it does show anti-correlation with $B$ and $|B_l|$, while correlation with $B_t$ at all the frequencies.

\section{Summary and Discussion}
We have analyzed high-resolution G~band and Ca~{\sc ii}~H  line filtergrams of the active region NOAA 10953 obtained by {\it Hinode}/SOT along with near simultaneous spectro-polarimetric observations of this active region from {\it Hinode} SOT/SP to study the relation between various magnetic parameters and the intensity oscillatory power in the photosphere and chromosphere. Our chief findings are as follows:

\begin{enumerate}
\item Photospheric power maps derived from G~band time series reveal more power in the quiet Sun (weaker fields) in comparison to the sunspot. This is in agreement with the previous reports that stronger magnetic fields absorb more power in the 5-minute band (Braun~et~al. \cite{Braun1992}; Hindman \& Brown \cite{Hindman1998}; Kumar~et~al. \cite{Kumar2000}; Venkatakrishnan et~al. \cite[and references therein]{PVK2002}). We, however, do not observe a reversal of this trend at higher frequencies in magnetic concentrations as shown by Venkatakrishnan et~al. (\cite{PVK2002}) in oscillatory power derived from photospheric Dopplergrams. Our analysis showing the absence of photospheric power at high frequencies in strong magnetic field regions is consistent with the results of Jain \& Haber (\cite{Jain2002}) who reported that only power spectra derived from Dopplergrams exhibit the above trend and not from intensity filtergrams. We emphasize here that the well known 5-minute oscillations are not dominantly seen in G~band power maps, whereas the 5-minute oscillations have been distinctly observed by Jain \& Haber (\cite{Jain2002}) in the continuum intensity power of the data obtained from {\it SoHO}/MDI. 

To confirm this, we have analyzed Dopplergrams and continuum intensity images obtained by the Helioseismic and Magnetic Imager (HMI) instrument onboard {\it Solar Dynamics Observatory} ({\it SDO}) on 2013 April 11. We estimated the PSD of the quiet Sun near the disk center for both the data sets. The Dopplergrams exhibit dominant power in 5-minute regime and similar nature of power is also observed in the continuum intensity images. However, the power in intensity is weaker in comparison to the Doppler power in the 5-minute regime. These results obtained with {\it SDO}/HMI is in agreement with the work done by Jain \& Haber (\cite{Jain2002}) using the data obtained from {\it SoHO}/MDI. We conjecture that the reason behind the observed lower power of 5-minute oscillations in G~band data relative to that in continuum intensity power from {\it SoHO}/MDI and {\it SDO}/HMI could be due to difference in their formation heights in the solar atmosphere. 

\item It is well known that in the chromosphere, the umbra shows an enhancement in power at frequencies above 5~mHz (Bhatnagar \& Tanaka \cite{Bhatnagar1972}; Braun~et~al. \cite{Braun1992}; Brown~et~al. \cite{Brown1992}; Kentischer \& Mattig \cite{Kenti1995}; Lites \cite{Lites1986}). We also observe such a behavior in our analysis of  Ca~{\sc ii}~H power (c.f., Figure~3 and 5). The umbral 3-minute chromospheric oscillations are suggested to emanate directly from the photosphere through linear wave propagation (Centeno et~al. \cite{Centeno2006}). Spectropolarimetric investigations by Centeno et~al. (\cite{Centeno2006}) using simultaneous observations taken in photosphere and chromosphere have provided observational evidence for the upward propagation of slow magneto-acoustic waves from the photosphere to the chromosphere inside the umbra of a sunspot. The phase spectra derived by Centeno et~al. (\cite{Centeno2006}) yield a value of the atmospheric cut-off frequency around 4~mHz and shows evidence for the upward propagation of higher frequency oscillations. Similarly, the presence of 5-minute oscillations in the chromospheric quiet Sun, and penumbral regions is attributed to the inclined magnetic field lines along which the photospheric 5-minute oscillations propagate to the higher atmosphere (Jefferies~et~al.~\cite{Jeff2006}; McIntosh \& Jefferies~\cite{McI2006}).

\item We have used near simultaneous {\it Hinode} SOT/SP observations of an active region, close to disk-center, to study the relationship between the magnetic-field parameters and the intensity oscillatory power in G~band and Ca~{\sc ii}~H. The azimuthally averaged radial profiles of field strength and inclination show a smooth decrease from the centre of the umbra to the periphery of the sunspot (c.f., Figure~\ref{SPvar}). However, the power does not exhibit a similar behavior. Umbra in Ca~{\sc ii}~H and G~band shows reduction of power in 0-2~mHz band. While this behavior is illustrated at all other frequencies in G~band, the same shows enhancement in 5-8~mHz band in Ca~{\sc ii}~H. The umbra-penumbra boundary shows enhancement of G~band power at all frequencies, where the transverse magnetic field is the highest. 
  
\item  The correlation analysis to the umbra and penumbra, separately, shows correlation between power and magnetic field parameters. In order to examine the influence of magnetic field on oscillatory power, we determined the correlation as a function of frequency bins. We observe that the photospheric magnetic parameters (except the transverse component) are correlated with umbral power in Ca~{\sc ii}~H and anti-correlated with umbral power in G~band at all frequencies, while the transverse magnetic field exhibits the opposite result. However, the analysis that includes only the penumbra shows that G~band power in the penumbra is anti-correlated with magnetic field parameters at all frequencies. Ca~{\sc ii}~H power in penumbra shows strong correlation with photospheric magnetic field in the frequency band 3.0-3.5~mHz, while at every other frequencies they are anti-correlated. 

The observed correlation between chromospheric penumbral power and photospheric magnetic fields in the 5-minute band could be the result of the magnetic inclination becoming large enough to allow photospheric 5-minute power to tunnel through higher acoustic cut-off frequency as demonstrated by Bloomfield et~al. (\cite{Bloom2007}) using simultaneous spectro-polarimetric observations of photosphere and chromosphere. As the umbral 3-minute chromospheric oscillations are inferred as field-aligned propagating slow magneto-acoustic waves (Centeno et~al. \cite{Centeno2006}), the observed correlation between photospheric magnetic fields and the chromospheric umbral power could also be due to the transportation of the photospheric power to the chromosphere through the magnetic field lines.

\end{enumerate}

\begin{acknowledgements}
{\it Hinode} is a Japanese mission developed and launched by ISAS/JAXA, collaborating with NAOJ as a domestic partner, NASA and STFC (UK) as international partners. Scientific operation of the {\it Hinode} mission is conducted by the {\it Hinode} science team organized at ISAS/JAXA. This team mainly consists of scientists from institutes in the partner countries. Support for the post-launch operation is provided by JAXA and NAOJ (Japan), STFC (U.K.), NASA (U.S.A.), ESA, and NSC (Norway). We also thank Community Spectro-polarimetric Analysis Center team at High-Altitude Observatory for providing the vector magnetogram (level2) data. Thanks to B.~Ravindra for his valuable suggestions related to this work. Rohan E. Louis is grateful for the financial assistance from the German Science Foundation (DFG) under grant DE 787/3-1. We thank the anonymous referee for constructive comments and suggestion that improved the presentation of the manuscript.

\end{acknowledgements}

\end{document}